\documentstyle[preprint,aps,tighten]{revtex}
\begin{document}
\draft

\title{Sensitivity of Quantum Chaotic Wavefunction Intensities to
Changes in External Perturbations}
\author{E. R. Mucciolo$^1$, B. D. Simons$^2$, A. V. Andreev$^{3}$,
and V. N. Prigodin$^4$}
\address
{$^1$Nordita, Blegdamsvej 17, DK-2100 Copenhagen {\O}, Denmark\\
$^2$Blackett Laboratory, Imperial College, London, SW7\ 2BZ, UK\\
$^3$Department of Physics, Massachusetts Institute of Technology, 77
Massachusetts Avenue, Cambridge, MA 02139\\
$^4$Max-Planck-Institute f\"ur Physik Komplexer Systeme,
Au\ss enstelle Stuttgart, Heisenbergstr. 1, 70569 Stuttgart, Germany}
\maketitle
\begin{abstract}
We examine the sensitivity of wavefunction intensities in chaotic
quantum systems to small changes in an arbitrary external perturbation.
A universal scaling is proposed for all three Dyson ensembles and a
novel theoretical approach is used to determine exact expressions for
systems which violate T-invariance. Analytical results are compared
with numerical simulations of tight-binding Anderson Hamiltonians.
\end{abstract}
\par

Systems which are classically chaotic show a wide degree of
universality in their quantum properties \cite{Berry,chaos}. The
generality of this description unites a large class of systems which
include examples as diverse as the spectra of atomic nuclei
\cite{Porter65}, Rydberg atoms in strong magnetic fields
\cite{Delande91}, weakly disordered metallic grains
\cite{Gorkov65,Efetov83}, and ballistic electronic cavities
\cite{Marcus93}. Studies have focussed largely on the statistical
properties of spectra and wavefunction intensities. Recently, there
has been an attempt to extend the universal description to encompass
the response of spectra to changes in an arbitrary external
perturbation \cite{correlations,Altshuler95}. In this letter we will
show how this description extends to the study of parametric
correlations in properties of wavefunctions amplitudes.

Traditional studies have involved defining spectral averages at fixed
values of external fields \cite{Mehta91}. From an experimental point
of view it is clearly more desirable to understand what happens to the
statistics of a given eigenfunction at a certain position upon varying
an external parameter such as magnetic field. Such information is
contained within the joint distribution function
\begin{eqnarray}
W({\bar v},v;X) & = & \Big\langle\delta\biglb({\bar v}+v/2-
\Omega|\psi_{\nu}({\vec r}, {\bar X}+X/2)|^2\bigrb) \nonumber \\
& & \times \delta\biglb({\bar v}-v/2-\Omega|\psi
_{\nu}({\vec r},{\bar X} -X/2)|^2\bigrb)\Big\rangle,
\label{eq:wdef}
\end{eqnarray}
where $\psi_{\nu}({\vec r},X)$ represents the eigenfunction of the
$\nu$-th state at a position ${\vec r}$, say inside a quantum dot, $X$
parametrizes some external perturbation such as magnetic field, and
$\Omega$ denotes the total volume of the system. Statistical
averaging, denoted by $\langle\cdots\rangle$ can in principle be
performed over some range of energy, parameter, or over the
statistical ensemble if one is defined. Our study is motivated by a
number of recent experiments that, in principle, allow $W$ to be
measured directly. These include STM measurements in nanostructures or
``quantum corrals'' \cite{STM}, microwave cavities \cite{muwave}, and
small acoustic resonators \cite{acoustics}.

The universality of the statistical properties of weakly disordered
metallic grains has been verified through the supersymmetry approach
\cite{Efetov83}. On time scales longer than the typical diffusion time
$t_D=L^2/D$, where $D=v_F^2 \tau/d$ denotes the classical diffusion
coefficient, statistical properties of the spectra and wavefunctions
coincide with those of random matrix ensembles. In this regime there
exist only two relevant parameters, the average level spacing
$\Delta=\langle E_{\nu+1}-E_\nu\rangle$, and the mean square level
gradient $C(0) \equiv\langle (\partial \epsilon_\nu/ \partial
X)^2\rangle$ (typically $C(0)\sim 1/t_D\Delta$). The dependence of
statistical properties on both can be removed by a straightforward
unfolding of the energy levels $\epsilon_\mu=E_\mu/\Delta$, and a
rescaling of the perturbation $x = X \sqrt{C(0)}$. Strong numerical
evidence (for a review, see Refs.~\cite{Altshuler95,Bohigas91})
suggests that this type of universality extends to all non-integrable
systems and serves as a characterization of the general phenomenum of
quantum chaos.

Below we will show that, asymptotically in the limit $x\to 0$, the
joint distribution function takes the universal form
\begin{equation}
W({\bar v},v;x) = P_\beta ({\bar v})\ \frac{1}{2\pi x\sqrt{\bar v}}
\ B_\beta\bigg(\frac{v}{2\pi x\sqrt{\bar v}}\bigg),
\label{eq:wans}
\end{equation}
where
$P_\beta(\bar{v})=(\bar{v}\beta/2)^{\beta/2-1}e^{-\bar{v}\beta/2}\
(\beta/2)/\Gamma(\beta/2)$ denotes the $\chi^2$ distribution of
wavefunction intensities \cite{Porter65} for Dyson ensembles of
Hermitian random matrices, and $B_\beta(s)$ is a universal function
whose form depends only on the symmetries of the system. By
convention, T-invariant ensembles that conserve/violate spin
rotational symmetry are termed orthogonal ($\beta=1$)/symplectic
($\beta=4$) respectively. Those which violate T-invariance are known
as unitary $(\beta=2)$. For the latter, a new approach to calculate
the statistical properties of spectral determinants \cite{Andreev95}
yields the following expression for the correlator
\begin{equation}
B_2(s)=\frac{35+14s^2+3s^4}{12\pi(1+s^2)^4}.
\label{eq:bs}
\end{equation}
The powerlaw behavior of $s\gg1$, which implies a large sensitivity,
is the related to the algebraic level repulsion of the Wigner-Dyson
statistics.

Before comparing $W$ with numerical simulations, we will show how
random matrix theory (RMT) leads to the general scaling in
Eq.~(\ref{eq:wans}). We emphasize that the same conclusions could be
drawn from the study of a microscopic theory involving weakly
disordered metallic grains. Consider the Hamiltonian
\begin{equation}
H=H_0+X H_1,
\end{equation}
where $H_0$ and $H_1$ denote $N\times N$ complex random Hermitian
matrices which belong to the Gaussian unitary ensemble defined by the
distribution $P(H)\ dH\propto\exp[-(N/2\lambda^2){\rm Tr} H^2]\ dH$.

For small values of $X$ the distribution $W$ involves two main types
of contributions. The first arises when the change in the perturbation
induces an anticrossing of the level $\nu$ and can not be treated
within perturbation theory. However, since it scales in proportion to
the number of anticrossings encountered it vanishes as
$X^{\beta+1}$. The remaining contribution can be treated
perturbatively in $X$. We will be concerned with the leading order of
perturbation theory which becomes asymptotically exact in the limit
$X\to 0$. The numerical simulations below show that even for small
non-zero values of $X$ this contribution gives an excellent
approximation to $W$. To the first order of perturbation theory, an
expansion around $X$ gives (${\vec r}\to \alpha$)
\begin{equation}
W({\bar v},v;X)=\bigg\langle \delta\biglb({\bar v}- A_\nu({\bar
X})\bigrb)\ \delta\bigglb(v-X\frac{\partial A_\nu}
{\partial\bar{X}}({\bar X})\biggrb) \bigg\rangle,
\label{eq:wdef2}
\end{equation}
where
\begin{equation}
\frac{\partial A_\nu}{\partial\bar{X}}({\bar X}) = \Omega
\sum_{\mu\ne \nu}\bigg[{\psi_\mu(\alpha;{\bar X})[H_1]_{\mu\nu}
\psi_\nu^{\ast}(\alpha;{\bar X})\over E_\nu({\bar X})-
E_\mu({\bar X})}+{\rm c.c.}\bigg],
\label{eq:wdef3}
\end{equation}
and $A_\nu({\bar X})=\Omega |\psi_\nu(\alpha;{\bar X})|^2$ denotes the
intensity of the wavefunction at site $\alpha$. Later we will discuss
the role of the higher order contributions in controlling the
asymptotic form of $W$ and restricting the region over which
Eq.~(\ref{eq:wans}) is applicable.

After Fourier transforming, and averaging over the Gaussian
perturbation $H_1$, we obtain
\begin{mathletters}
\begin{eqnarray}
W({\bar v},v;x) & = & \int_{-\infty}^{\infty} {dt\over 2\pi}\
e^{-ivt}\ {\widetilde W}({\bar v},t;X), \\
{\widetilde W}({\bar v},t;x) & = &\bigg\langle\delta({\bar v}-A_\nu)
\exp\bigg[ -t^2 x^2 \sum_{\mu\ne\nu} {A_\nu A_\mu\over
(\epsilon_\mu-\epsilon_\nu)^2}\bigg]\bigg\rangle,
\end{eqnarray}
\end{mathletters}
where we have made use of the rescaling with
$C(0)=\lambda^2/N\Delta^2$.

Existing studies, such as the supersymmetry approach, show that, in
the universal (long-time) limit, the statistical properties of the
wavefunctions become decoupled from the spectrum \cite{Prigodin95}.
As a result, the former can be averaged separately and we obtain
Eq.~(\ref{eq:wans}), with
\begin{mathletters}
\begin{eqnarray}
B_\beta(s) & = & \int_{-\infty}^{\infty}\frac{dk}{2\pi}\ e^{-isk}
{\widetilde B}_\beta(k), \\
{\widetilde B}_\beta(k) & = &
\left\langle\prod_{\mu\ne\nu}\bigg[1+\frac{(2/\beta)(k/2\pi)^2}
{(\epsilon_\mu-\epsilon_\nu)^2}\bigg]^{-{\beta/2}}\right\rangle.
\label{eq:bsdef}
\end{eqnarray}
\end{mathletters}

The remaining average in Eq.~(\ref{eq:bsdef}) is performed over the
Wigner-Dyson distribution of the $N$ eigenvalues $\{ \epsilon_\mu\} $.

The eigenvalue $\epsilon_\nu$ is identified by the correlator as
special. In the limit $N\to \infty$ the translational invariance of
the average allows us to set this eigenvalue to zero. This introduces
an additional factor of $|\det H^\beta|$ into the average over the
remaining $N-1$ eigenvalues, and leads to the expression
\begin{equation}
{\widetilde B}_\beta(k) = \frac{1}{Z_{N\beta}}\left\langle
\frac{\det H^{2\beta}}{\det[H^2+(2/\beta)(k\Delta/2\pi)^2]^{\beta/2}}
\right\rangle,
\label{eq:wb}
\end{equation}
where $Z_{N\beta}= \langle |\det H^\beta|\rangle$ denotes a
normalization constant.

For unitary ensembles, an expression for Eq.~(\ref{eq:wb}) can be
found explicitly \cite{Andreev95}. The method, which is a
generalization of an approach introduced by Guhr \cite{Guhr93}, relies
on a superalgebraic construction in which the spectral determinant is
expressed as a Gaussian integral over fermion and bosonic variables.
By integrating out the original degrees of freedom involving the
matrix Hamiltonian $H$, an expression is found for ${\widetilde
B}_\beta(k)$ which involves only a $6\times 6$ supermatrix. In
contrast to the supersymmetry approach, instead of deriving a
non-linear $\sigma$-model from the effective theory, we make use of a
generalization of the Itzykson-Zuber integral \cite{Itzykson80} over a
pseudounitary group to obtain the expression shown in
Eq.~(\ref{eq:bs}).

Let us now consider the asymptotics of the distribution function.
Eqs.~(\ref{eq:wdef2}) and (\ref{eq:wdef3}) shows that large
fluctuations arise when levels become anomalously close. In this limit
$v\sim(x\sqrt{\bar{v}}/\omega)$, where
$\omega=|\epsilon_\nu-\epsilon_{\nu\pm 1}|\ll 1$ denotes the smallest
separation of two neighboring levels. RMT implies a probability
distribution for level separations of $p(\omega)\sim
\omega^\beta$~\cite{Mehta91}, which suggests
\begin{eqnarray}
W({\bar v},v;x)& \sim & P_\beta(\bar{v})\ p(\omega){d\omega\over dv}
\sim P_\beta(\bar{v})\frac{1}{x\sqrt{\bar{v}}}\ s^{-(\beta+2)},
\label{eq:asymp}
\end{eqnarray}
with $s=v/(2\pi x\sqrt{\bar{v}})$. This agrees with the large $s$
asymptotics in Eq.~(\ref{eq:bs}). As $s$ increases the perturbative
expansion eventually fails. In particular, for $\bar{v}>1$ higher
orders of perturbation theory in $x$ become important when $s\sim
1/x$, while for $\bar{v}<1$ non-perturbative contributions begin to
dominate at $s\sim\sqrt{\bar{v}}/x$. At the end-points $|v|=2\bar{v}$
the distribution is essentially controlled by level anticrossings and
Eq.~(\ref{eq:wans}) does not apply. One can show that
$W_\beta(\bar{v},v\rightarrow\pm 2\bar{v};x)\sim x^{\beta+1}|v\mp
2\bar{v}|^{\beta/2-1}P_\beta(2\bar{v})$.

The most striking feature of these results is the powerlaw decay of
the correlator $B_\beta(s)$, in contrast to the usual exponential
behavior of other types of wavefunction distributions
\cite{Prigodin95}. This admits to the existence of large fluctuations
of the wavefunction even for small perturbations. The typical scale of
the relaxation increases is proportion to $x$ and $\sqrt {\bar v}$.

To verify the validity of the theoretical predictions, we have
compared Eqs.~(\ref{eq:wans}) and (\ref{eq:bs}) with data taken from
numerical simulations for the three symmetry classes. Each ensemble
was defined by 45 realizations of a $7\times 7\times 7$ tight-binding
Anderson Hamiltonian. The diffusive behaviour of the majority of
states was ensured by choosing the ``on-site'' disorder potential
randomly from the range $[-3,3]$, and a nearest-neighbor hopping
integral normalized to unity.

For the unitary case, we adopted the geometry of a torus for the
lattice and an Aharonov-Bohm (A-B) flux was employed as the
perturbation, $X$. To ensure unitary symmetry for even vanishing flux,
fixed quasi-periodic boundary conditions were applied in the
transverse directions. In both the orthogonal and symplectic cases,
the boundary conditions in one direction were left open creating a
lattice geometry of a ring. In both cases, a ``radial'' electric field
served as the external parameter, $X$. Symplectic symmetry was
imposed by spin-dependent hopping elements mimicing the action of a
spin-orbit coupling.

The numerical results were accumulated by taking an averaging over a
wide range of values of the external perturbation, $X$. To minimize
the systematic variation in $\Delta$ and $C(0)$ as a function of
energy, only a fifth of the total number states, located near the
center of the band, were selected.

The results for each Dyson ensemble are shown in Figs.~\ref{fig:fig1},
\ref{fig:fig2}, and \ref{fig:fig3}. In all cases, the factorization
and scaling of Eq.~(\ref{eq:wans}) is born out. For $v=0$, good
quantitative agreement with the results implied by the $\chi^2$
distribution is found for all three ensembles. Moreover, the results
in Fig.~\ref{fig:fig1}(a) give excellent agreement with the analytical
expression in Eq.~(\ref{eq:bs}). For the latter, the deviation from
the theoretical prediction anticipated as a breakdown of perturbation
theory, seems to lie beyond the statistical accuracy of this numerical
work. Although we presently lack a complete analytical expression for
$B_1(s)$ and $B_4(s)$, the numerical data are consistent with the
powerlaw decay predicted in Eq.~(\ref{eq:asymp}).

In this letter we have examined the sensitivity of wavefunction
intensities in quantum chaotic systems to changes in an external
perturbation. The rescaling implied by Eq.~(\ref{eq:wans}) has been
verified numerically for all three symmetry classes and provides a
concrete prediction for experiment. An analytical expression has been
provided for unitary ensembles which shows a good quantitative
agreement with numerical simulations. Our results indicate that
variations of an external perturbation cause large wavefunction
amplitude fluctuations to decay only as a power law.

We wish to thank B. L. Altshuler, T. Guhr, C. H. Lewenkopf, F. von
Oppen, and N. Taniguchi for valuable discussions. (BDS) and (AVA) wish
to acknowledge the support of JSEP No. DAAL 03-89-0001, EC grant
No. SCC-CT 90-00s20, and the hospitality of NECI where some of this
work was performed. (BDS) is also greatful to the support of the Royal
Society.

\begin{figure}
\caption{Comparison of the theoretical estimate of the distribution
function $W(\bar{v},v;x)$ with results taken from 45 realizations of a
$7\times 7\times 7$ Anderson model with an A-B flux (unitary
ensemble). (a) shows the variation of $W(\bar{v},v;x)$ with $v$ for
fixed values of $x$ and $\bar{v}$; (b) shows $x\times W({\bar v},0;x)$
for different values of $x$. The solid lines are the theoretical
predictions. The inset shows the distribution of wave function
intensities compared to the $\chi^2$ distribution $P_2(s)$.}
\label{fig:fig1}
\end{figure}

\begin{figure}
\caption{Same as in Fig.~1, but without A-B flux or spin-orbit
coupling (orthogonal ensemble). The dashed lines in (a) and (b) are
the result of fittings based on Eqs. (10) and (2), respectively. The
inset shows the distribution of wavefunction intensities compared to
$P_1(s)$.}
\label{fig:fig2}
\end{figure}

\begin{figure}
\caption{As in Fig.~2, but now including a spin-dependent hopping
term (symplectic ensemble). The inset shows the distribution of
wavefunction intensities compared to $P_4(s)$.}
\label{fig:fig3}
\end{figure}

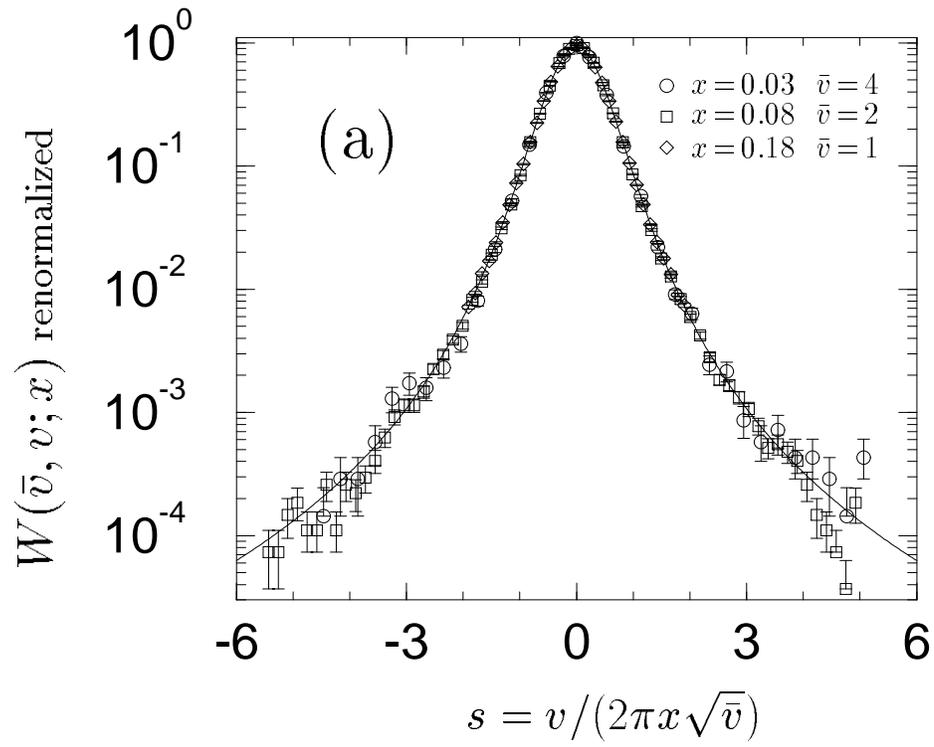
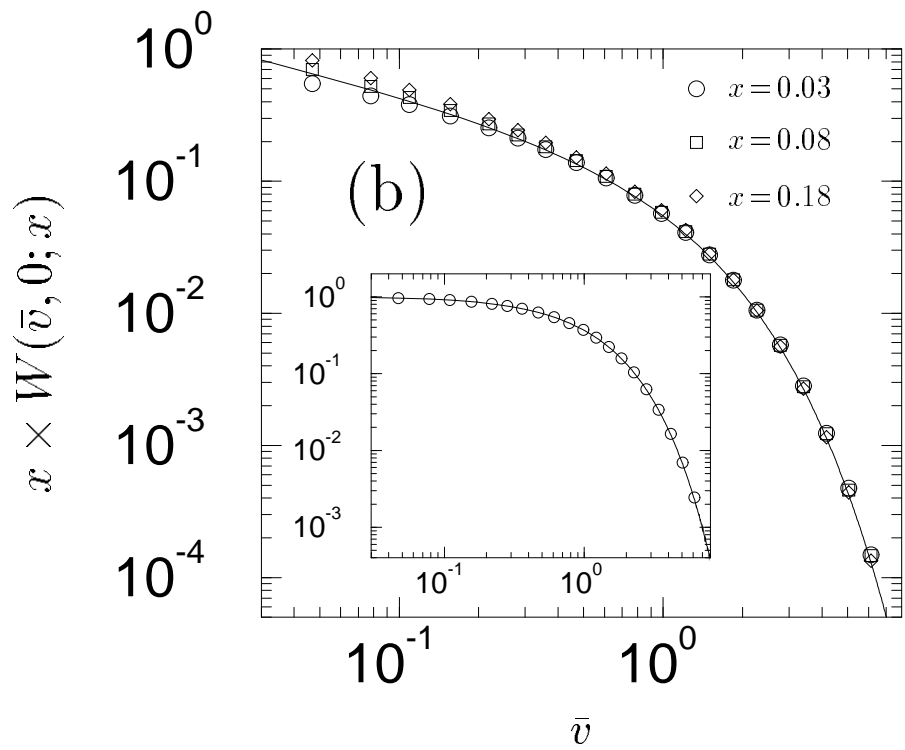
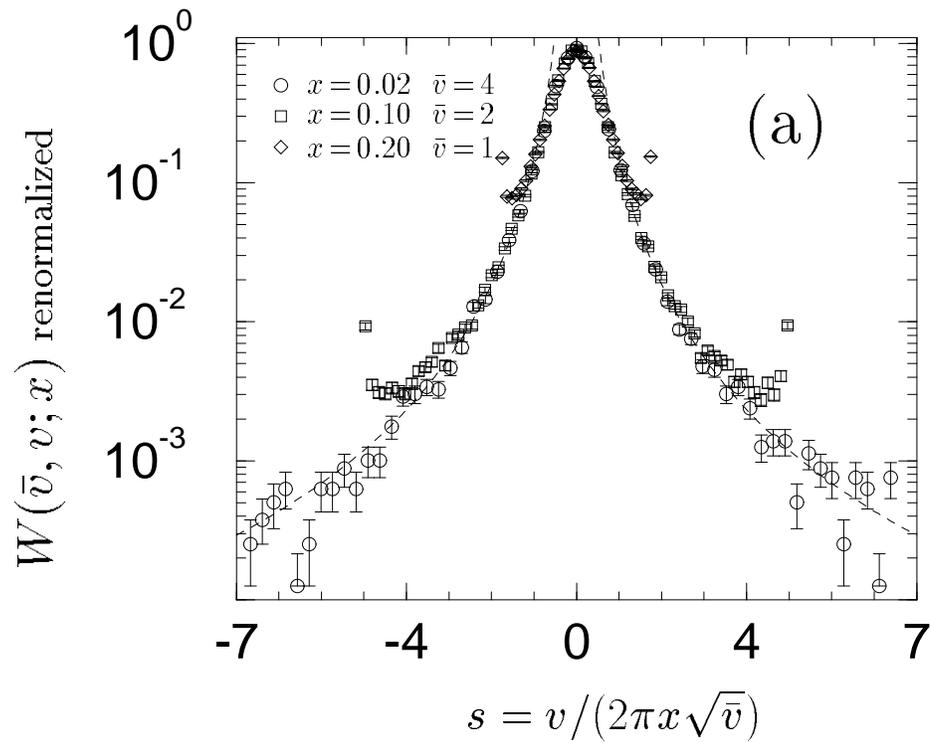
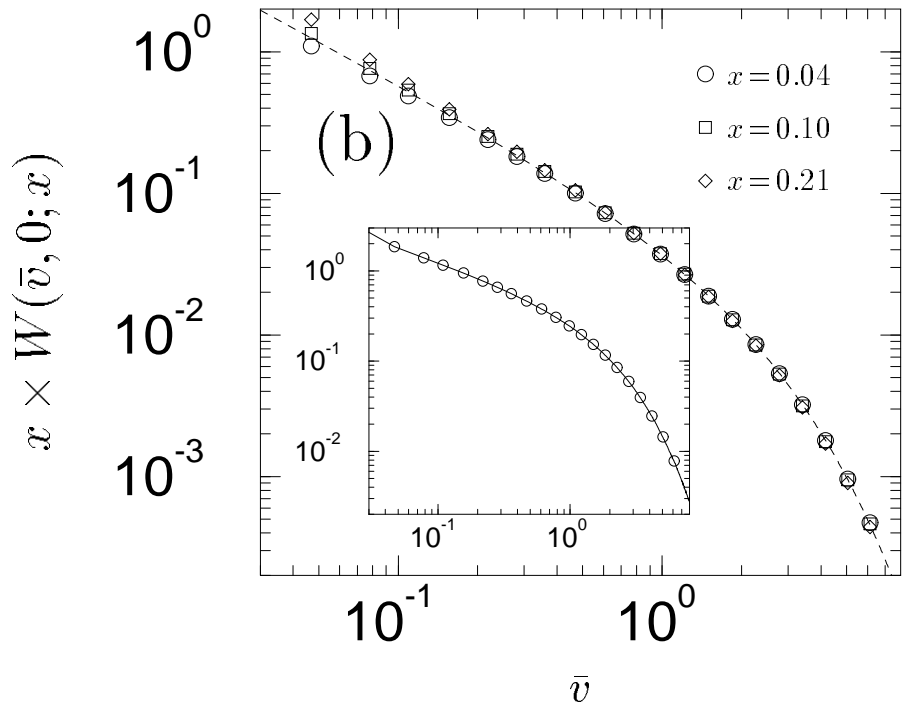
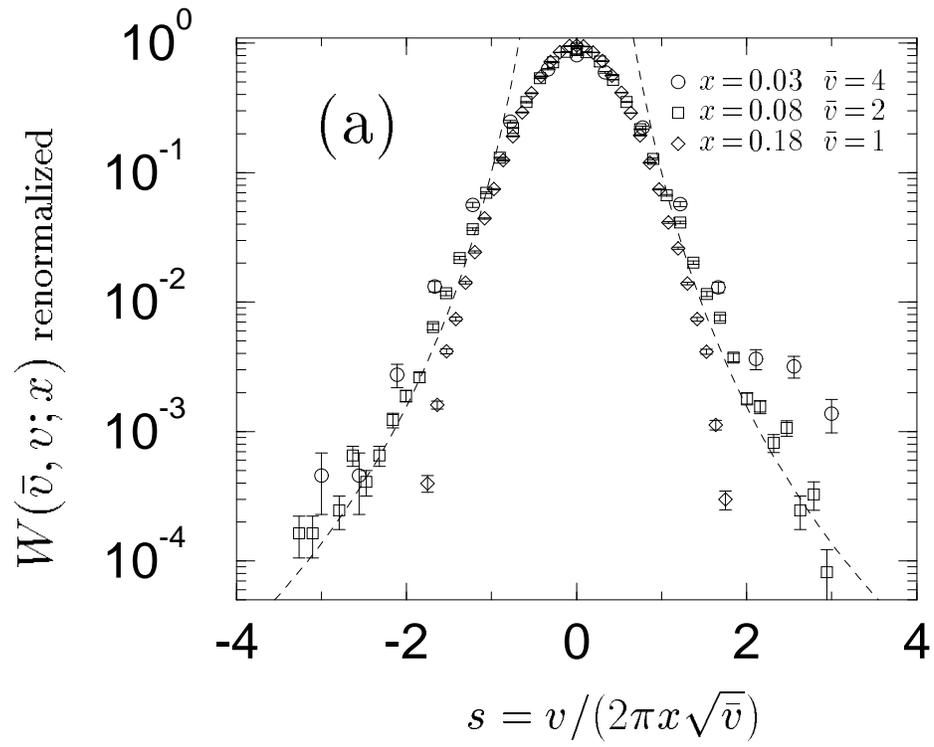
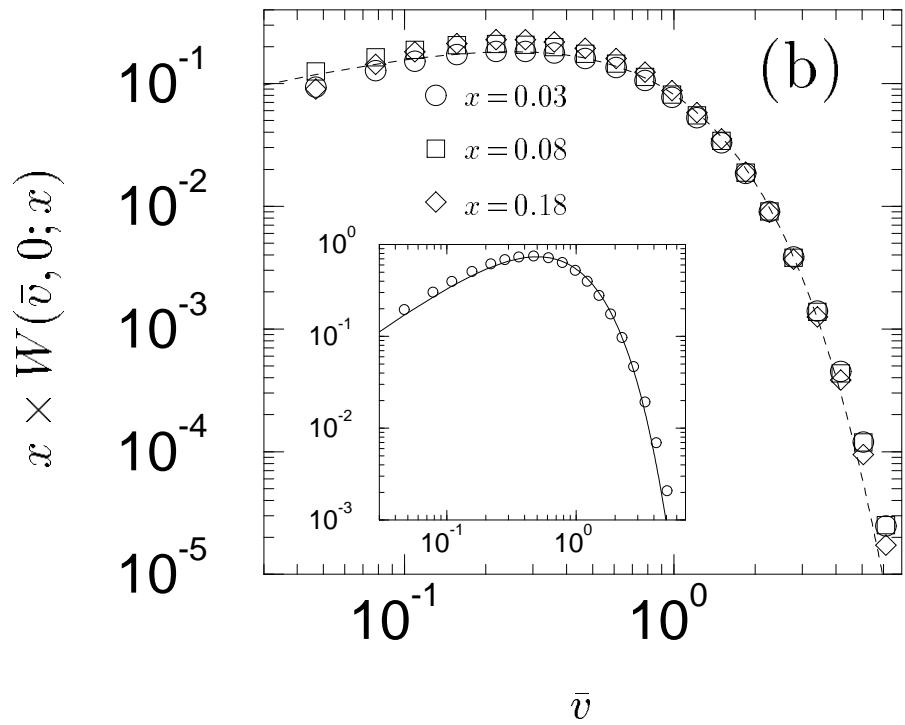
\end{document}